\documentclass[10pt]{article}  % -*- coding: utf-8 -*-

\usepackage{amsmath}
\usepackage{amssymb}
\usepackage{graphicx}
\usepackage{cite}
\usepackage{color}

\usepackage[labelfont=bf,labelsep=period,justification=raggedright]{caption}

\makeatletter
\renewcommand{\@biblabel}[1]{\quad#1.}
\makeatother

\date{}

\usepackage{etoolbox}  % patching other commands
\usepackage{outlines}
\usepackage{url}       % format and line-break URLs
\usepackage[utf8x]{inputenc}
\makeatletter
\@namedef{opt@inputenc.sty}{utf8}
\makeatother
\newcommand{\etal}{et~al.}
\newcommand{\inhead}[1]{\textbf{#1}}

\newcommand{\vocab}[1]{\emph{#1}}  % \emph toggles italics

\newcommand{\fixmecolor}{blue}

\let\oldoutline\outline
\def\outline{\oldoutline\color{\fixmecolor}}
\makeatletter
\patchcmd{\@make@cite@list}{\bfseries ?}{\textcolor{\fixmecolor}{\textbf{cite?}}}{}{}
\patchcmd{\@setref}{\bfseries ??}{\textcolor{\fixmecolor}{\textbf{ref?}}}{}{}
\makeatother

\graphicspath{{fig/}}
\begin{document}

\begin{flushleft}

% This title directly references the Ginsberg paper.
% (150 character limit; need also 50 character short version.)
{\Large\textbf{Global disease monitoring and forecasting with Wikipedia}}\\
Nicholas Generous,$^{\ast}$
Geoffrey Fairchild,
Alina Deshpande,
Sara Y.\ Del Valle,
Reid Priedhorsky
\\
\textbf{Defense Systems and Analysis Division, Los Alamos National Laboratory, Los Alamos, New Mexico, USA}
\\
$\ast$ Corresponding author: \url{generous@lanl.gov}

\end{flushleft}

\section{Abstract}

% 250 to 300 words recommended by PLOS

% Reid's formula:
% 1. Problem (what is is, why it's important)
% 2. Methods (what we did)
% 3. Results (what we learned)
% 4. Impact (why what we learned matters)

% Sara's formula:
% 1. The problem being addressed is...
% 2. This is important because...
% 3. The bottleneck that has kept the problem from being solved is...
% 4. The objective of this research is...
% 5. Results...
% 6. Policy recommendations...

% PLOS Medicine formula:
% 1. Background
% 2. Methods and findings
% 3. Conclusions

Infectious disease is a leading threat to public health, economic stability, and other key social structures. Efforts to mitigate these impacts depend on accurate and timely monitoring to measure the risk and progress of disease. Traditional, biologically-focused monitoring techniques are accurate but costly and slow; in response, new techniques based on social internet data such as social media and search queries are emerging. These efforts are promising, but important challenges in the areas of scientific peer review, breadth of diseases and countries, and forecasting hamper their operational usefulness.

We examine a freely available, open data source for this use: access logs from the online encyclopedia Wikipedia. Using linear models, language as a proxy for location, and a systematic yet simple article selection procedure, we tested 14 location-disease combinations and demonstrate that these data feasibly support an approach that overcomes these challenges. Specifically, our proof-of-concept yields models with $r^2$ up to 0.92, forecasting value up to the 28 days tested, and several pairs of models similar enough to suggest that transferring models from one location to another without re-training is feasible.

Based on these preliminary results, we close with a research agenda designed to overcome these challenges and produce a disease monitoring and forecasting system that is significantly more effective, robust, and globally comprehensive than the current state of the art.

% This is actually a very interesting point, but one we haven't made any effort to cover elsewhere. --Reid 4/4/14.
%
% Furthermore, the technique can be generalized to estimate numerous time-series phenomena provided they leave timely evidence that is observable by and of interest to laypeople.

\section{Author Summary}

% Please keep the Author Summary between 150 and 200 words Use first person.
% PLoS ONE authors please skip this step. Author Summary not valid for PLoS
% ONE submissions.

Even in developed countries, infectious disease has significant impact; for example, flu seasons in the United States take between 3,000 and 49,000 lives. Disease surveillance, traditionally based on patient visits to health providers and laboratory tests, can reduce these impacts. Motivated by cost and timeliness, surveillance methods based on internet data have recently emerged but are not yet reliable for several reasons, including weak scientific peer review, breadth of diseases and countries covered, and underdeveloped forecasting capabilities.

We argue that these challenges can be overcome by using a freely available data source: aggregated access logs from the online encyclopedia Wikipedia. Using simple statistical techniques, our proof-of-concept experiments suggest that these data are effective for predicting the present as well as forecasting up to the 28-day limit of our tests. Our results also suggest that these models can be used even in places with no official data upon which to build models.

In short, this paper establishes the utility of Wikipedia as a broadly effective data source for disease information, and we outline a path to a reliable, scientifically sound, operational, global disease surveillance system that overcomes key gaps in existing traditional and internet-based techniques.

\section{Introduction}

\subsection{Motivation and overview}

Infectious disease remains extremely costly in both human and economic terms. For example, the majority of global child mortality is due to conditions such as acute respiratory infection, measles, diarrhea, malaria, and HIV/AIDS~\cite{lopez_global_2006}. Even in developed countries, infectious disease has great impact; for example, each influenza season costs the United States between 3,000 and 49,000 lives~\cite{thompson_estimates_2010} and an average of \$87~billion in reduced economic output~\cite{molinari_annual_2007}.

Effective and timely disease surveillance --- that is, detecting, characterizing, and quantifying the incidence of disease --- is a critical component of prevention and mitigation strategies that can save lives, reduce suffering, and minimize impact. Traditionally, such monitoring takes the form of patient interviews and/or laboratory tests followed by a bureaucratic reporting chain; while generally considered accurate, this process is costly and introduces a significant lag between observation and reporting.

These problems have motivated new surveillance techniques based upon internet data sources such as search queries and social media posts. Essentially, these methods use large-scale data mining techniques to identify health-related activity traces within the data streams, extract them, and transform them into some useful metric. The basic approach is to train a statistical estimation model against ground truth data, such as ministry of health disease incidence records, and then apply the model to generate estimates when the true data are not available, e.g., when forecasting or when the true data have not yet been published. This has proven effective and has spawned operational systems such as Google Flu Trends (\url{http://www.google.org/flutrends/}). However, four key challenges remain before internet-based disease surveillance models can be reliably integrated into an decision-making toolkit:

\begin{enumerate}

  % NOTE: If we had more freedom of packages, we could use the enumitem
  % package for custom labels and wouldn't need to make them explicit. That
  % would be better DRY.

\item[C1.] \inhead{Openness.} Models should afford review, replication, improvement, and deployment by third parties. This guarantees a high-quality scientific basis, continuity of operations, and broad applicability. These requirements imply that model algorithms --- in the form of source code, not research papers --- must be generally available, and they also imply that complete input data must be available. The latter is the key obstacle, as terms are dictated by the data owner rather than the data user; this motivated our exploration of Wikipedia access logs. To our knowledge, no models exist that use both open data and open algorithms.

\item[C2.] \inhead{Breadth.} Dozens of diseases in hundreds of countries have sufficient impact to merit surveillance; however, adapting a model from one \vocab{disease-location context} to another can be costly, and resources are often, if not usually, unavailable to do so. Thus, models should be cheaply adaptable, ideally by simply entering new incidence data for training. While most published models afford this flexibility in principle, few have been expressly tested for this purpose.

\item[C3.] \inhead{Transferability.} Many contexts have insufficient reliable incidence data to train a model (for example, the relevant ministry of health might not track the disease of interest), and in fact these are the contexts where new approaches are of the greatest urgency. Thus, trained models should be translatable to new contexts using alternate, non-incidence data such as a bilingual dictionary or census demographics. To our knowledge, no such models exist.

\item[C4.] \inhead{Forecasting.} Effective disease response depends not only on the current state of an outbreak but also its future course. That is, models should provide not only estimates of the current state of the world --- \vocab{nowcasts} --- but also \vocab{forecasts} of its future state.

While recent work in disease forecasting has made significant strides in accuracy, forecasting the future of an outbreak is still a complex affair that is sharply limited in contexts with insufficient data or insufficient understanding of the biological processes and parameters underpinning the outbreak. In these contexts, a simpler statistical approach based on leading indicators in internet data streams may improve forecast availability, quality, and time horizon. Prior evaluations of such approaches have yielded conflicting results and to our knowledge have not been performed at time granularity finer than one week.

\end{enumerate}

In order to address these challenges, we propose a new approach based on freely available Wikipedia article access logs. In the current proof of concept, we use language as a proxy for location, but we hope that access data explicitly aggregated by geography will become available in the future. (Our  implementation is available as open source software: \url{http://github.com/reidpr/quac}.) To demonstrate the feasibility of techniques built upon this data stream, we built linear models mapping daily access counts of encyclopedia articles to case counts for 7 diseases in 9 countries, for a total of 14 contexts. Even a simple article selection method was successful in 8 of the 14 contexts, yielding models with $r^2$ up to 0.89 in nowcasting and 0.92 in forecasting, with most of the successful contexts having forecast value up to the tested limit of 28 days. Specifically, we argue that approaches based on this data source can overcome the four challenges as follows:

\begin{enumerate}

\item[C1.] Anyone with relatively modest computing resources can download the complete Wikipedia dataset and keep it up to date. The data can also be freely shared with others.

\item[C2.] In cases where estimation is practical, our approach can be adapted to a new context by simply supplying a reliable incidence time series and selecting input articles. We demonstrate this by computing effective models for several different contexts even with a simple article selection procedure. Future, more powerful article selection procedures will increase the adaptability of the approach.

\item[C3.] In several instances, our models for the same disease in different locations are very similar; i.e., correlations between different language versions of the same article and the corresponding local disease incidence are similar. This suggests that simple techniques based on inter-language article mappings or other readily available data can be used to translate models from one context to another without re-training.

\item[C4.] Even our simple models show usefully high $r^2$ when forecasting a few days or weeks into the future. This suggests that the general approach can be used to build short-term forecasts with reasonably tight confidence intervals.

\end{enumerate}

In short, this paper offers two key arguments. First, we evaluate the potential of an emerging data source, Wikipedia access logs, for global disease surveillance and forecasting in more detail than is previously available, and we argue that the openness and other properties of these data have important scientific and operational benefits. Second, using simple proof-of-concept experiments, we demonstrate that statistical techniques effective for estimating disease incidence using previous internet data are likely to also be effective using Wikipedia access logs.

We turn next to a more thorough discussion of prior work, both to set the stage for the current work as well as outline in greater detail the state of the art's relationship to the challenges above. Following that, we cover our methods and data sources, results, and a discussion of implications and future work.

\subsection{Related work}

Our paper draws upon prior scholarly and practical work in three areas: traditional patient- and laboratory-based disease surveillance,  Wikipedia-based measurement of the real world, and internet-based disease surveillance.

\subsubsection{Traditional disease surveillance}

Traditional forms of disease surveillance are based upon direct patient contact or biological tests taking place in clinics, hospitals, and laboratories. The majority of current systems rely on syndromic surveillance data (i.e., about symptoms) including clinical diagnoses, chief complaints, school and work absenteeism, illness-related 911 calls, and emergency room admissions~\cite{kman_biosurveillance:_2012}.

For example, a well-established measure for influenza surveillance is the \vocab{fraction of patients with influenza-like illness}, abbreviated simply ILI. A network of outpatient providers report the total number of patients seen and the number who present with symptoms consistent with influenza that have no other identifiable cause~\cite{_overview_2012}. Similarly, other electronic resources have emerged, such as the Electronic Surveillance System for the Early Notification of Community Based Epidemics (ESSENCE), based on real-time data from the Department of Defense Military Health System~\cite{bravata_systematic_2004} and BioSense, based on data from the Department of Veterans Affairs, the Department of Defense, retail pharmacies, and Laboratory Corporation of America~\cite{borchardt_categorization_2006}. These systems are designed to facilitate early detection of disease outbreaks as well as response to harmful health effects, exposure to disease, or related hazardous conditions.

Clinical labs play a critical role in surveillance of infectious diseases. For example, the Laboratory Response Network (LRN), consisting of over 120 biological laboratories, provides active surveillance of a number of disease agents in humans ranging from mild (e.g., non-pathogenic \textit{E. coli} and \textit{Staphylococcus aureus}) to severe (e.g., Ebola and Marburg), based on clinical or environmental samples~\cite{kman_biosurveillance:_2012}. Other systems monitor non-traditional public health indicators such as school absenteeism rates, over-the-counter medication sales, 911 calls, veterinary data, and ambulance run data. For example, the Early Aberration Reporting System (EARS) provides national, state, and local health departments alternative detection approaches for syndromic surveillance~\cite{hutwagner_bioterrorism_2003}.

The main value of these systems is their accuracy. However, they have a number of disadvantages, notably cost and timeliness: for example, each ILI datum requires a practitioner visit, and ILI data are published only after a delay of 1--2 weeks~\cite{_overview_2012}.

\subsubsection{Wikipedia}

Wikipedia is an online encyclopedia that has, since its founding in 2001, grown to contain approximately 30 million articles in 287 languages~\cite{wikipedia_editors_wikipedia_2013}. In recent years, it has consistently ranked as a top-10 website; as of this writing, it is the 6th most visited website in the world and the most visited site that is not a search engine or social network~\cite{alexa_internet_inc._alexa_2013}, serving roughly 850 million article requests per day~\cite{wikimedia_foundation_page_2013}. For numerous search engine queries, a Wikipedia article is the top result.

Wikipedia contrasts with traditional encyclopedias on two key dimensions: it is free of charge to read, and anyone can make changes that are published immediately --- review is performed by the community \emph{after} publication. (This is true for the vast majority of articles. Particularly controversial articles, such as ``George W. Bush'' or ``Abortion'', have varying levels of edit protection.) While this surprising inversion of the traditional review-publish cycle would seem to invite all manner of abuse and misinformation, Wikipedia has developed effective measures to deal with these problems and is of similar accuracy to traditional encyclopedias such as \emph{Britannica}~\cite{giles_internet_2005}.

Wikipedia article access logs have been used for a modest variety of research. The most common application is detection and measurement of popular news topics or events~\cite{ahn_wikitopics:_2011,ciglan_wikipop:_2010,holaker_event_2013,osborne_bieber_2012,althoff_analysis_2013}. The data have also been used to study the dynamics of Wikipedia itself~\cite{priedhorsky_creating_2007,thij_modeling_2012,tran_cross_2013}. Social applications include evaluating toponym importance in order to make type size decisions for maps~\cite{burdziej_using_2012}, measuring the flow of concepts across the world~\cite{tinati_approach_2013}, and estimating the popularity of politicians and political parties~\cite{yasseri_can_2013}. Finally, economic applications include attempts to forecast movie ticket sales~\cite{mestyan_early_2013} and stock prices~\cite{moat_quantifying_2013}. The latter two applications are of particular interest because they include a forecasting component, as the present work does.

In the context of health information, the most prominent research direction focuses on assessing the quality of Wikipedia as a health information source for the public, e.g., with respect to
cancer~\cite{leithner_wikipedia_2010,rajagopalan_patient-oriented_2011},
carpal tunnel syndrome~\cite{lutsky_quality_2013},
drug information~\cite{kupferberg_accuracy_2011},
and kidney conditions~\cite{thomas_evaluation_2013}. To our knowledge, only four health studies exist that make use of Wikipedia access logs. Tausczik~\etal\ examined public ``anxiety and information seeking'' during the 2009 H1N1 pandemic, in part by measuring traffic to H1N1-related Wikipedia articles~\cite{tausczik_public_2012}. Laurent and Vickers evaluated Wikipedia article traffic for disease-related seasonality and in relation to news coverage of health issues, finding significant effects in both cases~\cite{laurent_seeking_2009}. Aitken~\etal\ found a correlation between drug sales and Wikipedia traffic for a selection of approximately 5,000 health-related articles~\cite{aitken_engaging_2014}. None of these propose a time-series model mapping article traffic to disease metrics.

The fourth study is a recent article by McIver and Brownstein, which uses statistical techniques to estimate the influenza rate in the United States from Wikipedia access logs~\cite{mciver_wikipedia_2014}. In the next section, we compare and contrast this article with the present work in the context of a broader discussion of such techniques.

In summary, use of Wikipedia access logs to measure real-world quantities is beginning to emerge, as is interest in Wikipedia for health purposes. However, to our knowledge, use of the encyclopedia for quantitative disease surveillance remains at the earliest stages.

\subsubsection{Internet-based disease surveillance}

Recently, new forms of surveillance based upon the social internet have emerged; these data streams are appealing in large part because of their real-time nature and the low cost of information extraction, properties complementary to traditional methods. The basic insight is that people leave traces of their online activity related to health observations, and these traces can be captured and used to derive actionable information. Two main classes of trace exist: \vocab{sharing} such as social media mentions of face mask use~\cite{mniszewski_understanding_2014} and \vocab{health-seeking behavior} such as Web searches for health-related topics~\cite{ginsberg_detecting_2008}. (In fact, there is evidence that the volume of internet-based health-seeking behavior dwarfs traditional avenues~\cite{rice_influences_2006,fox_online_2006}.)

In this section, we focus on the surveillance work most closely related to our efforts, specifically, that which uses existing single-source internet data feeds to estimate some scalar disease-related metric. For example, we exclude from detailed analysis work that
provides only alerts~\cite{collier_biocaster:_2008,zhou_monitoring_2013},
measures public perception of a disease~\cite{ritterman_using_2009},
includes disease dynamics in its model~\cite{shaman_real-time_2013},
evaluates a third-party method~\cite{olson_reassessing_2013},
uses non-single-source data feeds~\cite{collier_biocaster:_2008,freifeld_healthmap:_2008},
or crowd-sources health-related data (``participatory disease surveillance'')~\cite{chunara_flu_2013,chunara_online_2012}.
We also focus on work that estimates biologically-rooted metrics. For example, we exclude metrics based on seasonality~\cite{ayers_seasonality_2013,seifter_utility_2010} and over-the-counter drug sales volume, itself a proxy~\cite{lindh_head_2012}.

These activity traces are embedded in
search queries~\cite{  % 292 - 264 = 28 = 27 s + 1 sl
  althouse_prediction_2011,
  breyer_use_2011,               %
  carneiro_google_2009,
  chan_using_2011,
  cho_correlation_2013,          %
  cooper_cancer_2005,
  desai_norovirus_2012,
  dugas_influenza_2013,          %
  dukic_internet_2011,
  eysenbach_infodemiology:_2006,
  ginsberg_detecting_2008,
  hagihara_internet_2012,        %
  hulth_eye-opening_2010,
  hulth_web_2009,
  hulth_web_2011,
  jia-xing_gonorrhea_2013,       %
  kang_using_2013,               %
  ocampo_using_2013,
  pelat_more_2009,               %
  polgreen_using_2008,
  walcott_determination_2011,
  wilson_early_2009,
  xu_neural_2010,                %
  xu_predicting_2011,
  yang_association_2011,
  yuan_monitoring_2013,
  zheluk_internet_2013,
  zhou_notifiable_2010           %
},
social media messages~\cite{  % 310 - 294 = 16
  achrekar_predicting_2011,
  achrekar_twitter_2012,
  aramaki_twitter_2011,
  broniatowski_national_2013,
  chunara_social_2012,           %  5
  culotta_lightweight_2013,
  doan_enhancing_2012,
  gomide_dengue_2011,
  hirose_prediction_2012,
  kim_use_2013,                  % 10
  lamb_separating_2013,
  lampos_nowcasting_2012,
  lampos_tracking_2010,
  nagel_complex_2013,
  paul_you_2011,                 % 15
  signorini_use_2011
},
and web server access logs~\cite{
  johnson_analysis_2004,
  mciver_wikipedia_2014,
  xu_predicting_2011
}.
At a basic level, traces are extracted by counting query strings, words or phrases, or web page URLs that are related to the metric of interest, forming a time series of occurrences for each item. A statistical model is then created that maps these input time series to a time series estimating the metric's changing value. This model is trained on time period(s) when both the internet data and the true metric values are available and then applied to estimate the metric value over time period(s) when it is not available, i.e., \vocab{forecasting} the future, \vocab{nowcasting} the present, and \vocab{anti-forecasting} the past (the latter two being useful in cases where true metric availability lags real time).

Typically, this model is linear, e.g.:
\begin{equation}
M = \sum_{j=1}^J \alpha_j x_j
\end{equation}
where $x_j$ is the count of some item, $J$ is the total number of possible items (i.e., vocabulary size), $M$ is the estimated metric value, and $\alpha_j$ are selected by linear regression or similar methods. When appropriately trained, these methods can be quite accurate; for example, many of the cited models can produce near real-time estimates of case counts with correlations upwards of $r = 0.95$.

The collection of disease surveillance work cited above has estimated incidence for a wide variety of infectious and non-infectious conditions:
avian influenza~\cite{carneiro_google_2009},
cancer~\cite{cooper_cancer_2005},
chicken pox~\cite{pelat_more_2009},
cholera~\cite{chunara_social_2012},
dengue~\cite{althouse_prediction_2011,
             chan_using_2011,
             gomide_dengue_2011},
dysentery~\cite{zhou_notifiable_2010},
gastroenteritis~\cite{desai_norovirus_2012,
                      hulth_eye-opening_2010,
                      pelat_more_2009},
gonorrhea~\cite{jia-xing_gonorrhea_2013},
hand foot and mouth disease (HFMD)~\cite{xu_predicting_2011},
HIV/AIDS~\cite{zheluk_internet_2013,
               zhou_notifiable_2010},
influenza~\cite{achrekar_predicting_2011,  % 367 - 341 = 26
                achrekar_twitter_2012,
                aramaki_twitter_2011,
                broniatowski_national_2013,
                cho_correlation_2013,       % 5
                culotta_lightweight_2013,
                doan_enhancing_2012,
                dugas_influenza_2013,
                eysenbach_infodemiology:_2006,
                ginsberg_detecting_2008,    % 10
                hirose_prediction_2012,
                hulth_web_2009,
                hulth_web_2011,
                johnson_analysis_2004,
                kang_using_2013,            % 15
                kim_use_2013,
                lamb_separating_2013,
                lampos_nowcasting_2012,
                lampos_tracking_2010,
                mciver_wikipedia_2014,      %%%
                nagel_complex_2013,         % 20
                pelat_more_2009,
                paul_you_2011,
                polgreen_using_2008,
                signorini_use_2011,
                xu_neural_2010,             % 25
                yuan_monitoring_2013},
kidney stones~\cite{breyer_use_2011},
listeriosis~\cite{wilson_early_2009},
malaria~\cite{ocampo_using_2013},
methicillin-resistant \textit{Staphylococcus aureus} (MRSA)~\cite{dukic_internet_2011},
pertussis~\cite{nagel_complex_2013},
pneumonia~\cite{polgreen_using_2008},
respiratory syncytial virus (RSV)~\cite{carneiro_google_2009},
scarlet fever~\cite{zhou_notifiable_2010},
stroke~\cite{walcott_determination_2011},
suicide~\cite{hagihara_internet_2012,
              yang_association_2011},
tuberculosis~\cite{zhou_notifiable_2010},
and
West Nile virus~\cite{carneiro_google_2009}.

Closely related to the present work is an independent, simultaneous effort by McIver \& Brownstein to measure influenza in the United States using Wikipedia access logs~\cite{mciver_wikipedia_2014}. This study used Poisson models fitted with LASSO regression to estimate ILI over a 5-year period. The results, Pearson's $r$ of 0.94 to 0.99 against official data, depending on model variation, compare quite favorably to prior work that tries to replicate official data. More generally, this article's statistical methods are more sophisticated than those employed in the present study. However, we offer several key improvements:
\begin{itemize}

\item We evaluate 14 location-disease contexts around the globe, rather than just one. In doing so, we test the use of language as a location proxy, which was noted briefly as future work in McIver \& Brownstein. (However, as we detail below, we suspect this is not a reliable geo-location method for the long term.)

\item We test our models for forecasting value, which was again mentioned briefly as future work in McIver \& Brownstein.

\item We evaluate models for translatability from one location to another.

\item We present negative results and use these to begin exploring when internet-based disease surveillance methods might and might not work.

\item We offer a systematic, well-specified, and simple procedure to select articles for model inclusion.

\item We normalize article traffic by total language traffic rather than using a few specific articles as a proxy for it.

\item Our software is open source and has only freely available dependencies, while the McIver \& Brownstein code is not available and depends on proprietary components (Stata).

\end{itemize}

Finally, the goals of the two studies differ. McIver \& Brownstein wanted to ``develop a statistical model to provide near-time estimates of ILI activity in the US using freely available data gathered from the online encyclopedia Wikipedia''~\cite[p.\,2]{mciver_wikipedia_2014}. Our goals are to assess the applicability of these data to global disease surveillance for operational public health purposes and to lay out a research agenda for achieving this end.

These methods are the basis for at least one deployed, widely used surveillance system. Based upon search query data, Google Flu Trends offers near-real-time estimates of influenza activity in 29 countries across the world (15 at the province level); another facet of the same system, Google Dengue Trends (\url{http://www.google.org/denguetrends/}) estimates dengue activity in 9 countries (2 at the province level) in Asia and Latin America.

Having laid out the space of quantitative internet disease surveillance as it exists to the best of our knowledge, we now consider this prior work in the context of our four challenges:

\begin{enumerate}

\item[C1.] \inhead{Openness.}
Deep access to search queries from
Baidu, a Chinese-language search engine serving mostly the Chinese market (\url{http://www.baidu.com})%
~\cite{jia-xing_gonorrhea_2013,
       yuan_monitoring_2013,
       zhou_notifiable_2010};
Google~\cite{althouse_prediction_2011,  % 445 - 425 = 20
             breyer_use_2011,
             carneiro_google_2009,
             chan_using_2011,
             cho_correlation_2013,
             desai_norovirus_2012,
             dugas_influenza_2013,
             dukic_internet_2011,
             eysenbach_infodemiology:_2006,
             ginsberg_detecting_2008,
             hagihara_internet_2012,
             kang_using_2013,
             ocampo_using_2013,
             pelat_more_2009,
             walcott_determination_2011,
             wilson_early_2009,
             yang_association_2011,
             xu_neural_2010,
             xu_predicting_2011,
             zheluk_internet_2013};
Yahoo~\cite{cooper_cancer_2005,
            polgreen_using_2008};
and Yandex, a search engine serving mostly Russia and Slavic countries in Russian (\url{http://www.yandex.ru}), English (\url{http://www.yandex.com}), and Turkish (\url{http://www.yandex.com.tr})~\cite{zheluk_internet_2013},
as well as
purpose-built health website
search queries~\cite{hulth_eye-opening_2010,
                     hulth_web_2009,
                     hulth_web_2011}
and access logs~\cite{johnson_analysis_2004,
                      xu_predicting_2011}
are available only to those within the organizations, upon payment of an often-substantial fee, or by some other special arrangement. While tools such as Baidu Index (\url{http://index.baidu.com}), Google Trends (\url{http://www.google.com/trends/}), Google Correlate (\url{http://www.google.com/trends/correlate/}), and Yandex's WordStat (\url{http://wordstat.yandex.com}) provide a limited view into specific search queries and/or time periods, as do occasional lower-level data dumps offered for research, neither affords the large-scale, broad data analysis that drives the most effective models.

% Further, access is controlled by a single organization.

The situation is only somewhat better for surveillance efforts based upon
Twitter~\cite{achrekar_predicting_2011,   % 479 - 463 = 16
              achrekar_twitter_2012,
              aramaki_twitter_2011,
              broniatowski_national_2013,
              chunara_social_2012,        %  5
              culotta_lightweight_2013,
              doan_enhancing_2012,
              gomide_dengue_2011,
              hirose_prediction_2012,
              kim_use_2013,               % 10
              lamb_separating_2013,
              lampos_nowcasting_2012,
              lampos_tracking_2010,
              nagel_complex_2013,
              paul_you_2011,              % 15
              signorini_use_2011}.
While a small portion of the real-time message stream (1\%, or 10\% for certain grandfathered users) is available outside the company without substantial fees, terms of use prohibit sharing historical data needed for calibration between researchers. Access rules are similar or significantly more restrictive for alternative social media platforms such as Sina Weibo, the leading Chinese microblogging site (\url{http://weibo.com}), and Facebook. Consistent with this, we were able to find no research meeting our inclusion criteria based on either of these extremely popular systems.

We identified only one prior effort making use of open data, McIver \& Brownstein with Wikipedia access logs~\cite{mciver_wikipedia_2014}. Open algorithms in this field of inquiry are also very limited. Of the works cited above, again only one, Althouse \etal~\cite{althouse_prediction_2011}, claims general availability of their algorithms in the form of open source code.

Finally, we highlight the quite successful Google Flu and Dengue Trends as a case study in the problems of closed data and algorithms. First, because their data and algorithms are proprietary, there is little opportunity for the wider community of expertise to offer peer review or improvements (for example, the list of search terms used by Dengue Trends has never been published, even in summary form); the importance of these opportunities is highlighted by the system's well-publicized estimation failures during the 2012--2013 flu season~\cite{butler_when_2013} as well as more comprehensive scholarly criticisms~\cite{olson_reassessing_2013}. Second, only Google can choose the level of resources to spend on Trends; no one else, regardless of their available resources, can add new contexts or take on operational responsibility should Google choose to discontinue the project.

\item[C2.] \inhead{Breadth.} While in principle these surveillance approaches are highly generalizable, nearly all extant efforts address a small set of diseases in a small set of countries, without testing specific methods to expand these sets.

% e.g.: "this method may be extended to other countries and other diseases"~\cite{gomide_dengue_2011}

The key exception is Paul \& Dredze~\cite{paul_you_2011}, which proposes a content-based method, \vocab{ailment topic aspect model} (ATAM), to automatically discover a theoretically unbounded set of medical conditions mentioned in Twitter messages. This unsupervised machine learning algorithm, similarly to latent Dirichlet allocation (LDA)~\cite{blei_latent_2003}, accumulates co-occurring words into probabilistic \vocab{topics}. Lists of health-related lay keywords, as well as the text of health articles written for a lay audience, are used to ensure that the algorithm builds topics related to medical issues. A test of the method discovered 15 coherent condition topics including infectious diseases such as influenza, non-infectious diseases such as cancer, and non-specific conditions such as aches and pains. The influenza topic's time series correlated very well with ILI data in the United States.

However, we identify three drawbacks of this approach. First, significant curated text input data in the target language are required; second, output topics require expert interpretation; and third, the ATAM algorithm has several parameters that require expert tuning. That is, in order to adapt the algorithm to a new location and/or language, expertise in both machine learning as well as the target language are required.

In summary, to our knowledge, no disease measurement algorithms have been proposed that are extensible to new disease-location contexts solely by adding examples of desired output. We propose a path to such algorithms.

\item[C3.] \inhead{Transferability.} To our knowledge, no prior work offers trained models that can be translated from one context to another. We propose using the inter-language article links provided in Wikipedia to accomplish this translation.

\item[C4.] \inhead{Forecasting.} A substantial minority of the efforts in this space test some kind of forecasting method. (Note that many papers use the term \vocab{predict}, and some even misuse \vocab{forecast}, to indicate nowcasting.) In addition to forecasting models that incorporate disease dynamics (recall that these are out of scope for the current paper), two basic classes of forecasting exist: \vocab{lag analysis}, where the internet data are simply time-shifted in order to capture leading signals, and statistical forecast models such as linear regression.

% \2 Jia-xing~\cite{jia-xing_gonorrhea_2013}
%    \3 granularity: 1 month
%    \3 results: no leading effect
% \2 Johnson~\cite{johnson_analysis_2004}
%    \2 granularity: 1 week
%    \2 results: ``no consistent trend''
% \2 Pelat~\cite{pelat_more_2009}
%    \3 granularity: 1 week
%    \3 results: no leading effect

Lag analysis has shown mixed results in prior work. Johnson~\etal~\cite{johnson_analysis_2004}, Pelat~\etal~\cite{pelat_more_2009},
and Jia-xing~\etal~\cite{jia-xing_gonorrhea_2013} identified no reliable leading signals. On the other hand, Polgreen~\etal~\cite{polgreen_using_2008} used lag analysis with a shift granularity of one week to forecast positive influenza cultures as well as influenza and pneumonia mortality with a horizon of 5 weeks or more (though these indicators may trail the onset of symptoms significantly). Similarly, Xu~\etal~\cite{xu_predicting_2011} reported evidence that lag analysis may be able to forecast HFMD by up to two months, and Yang~\etal~\cite{yang_association_2011} used lag analysis with a granularity of one month to identify search queries that lead suicide incidence by up to two months.

The more complex method of statistical forecast models appears potentially fruitful as well. Dugas \etal\ tested several statistical methods using positive influenza tests and Google Flu Trends to make 1-week forecasts~\cite{dugas_influenza_2013}, and Kim~\etal\ used linear regression to forecast influenza on a horizon of 1~month~\cite{kim_use_2013}.

In summary, while forecasts based upon models that include disease dynamics are clearly useful, sometimes this is not possible because important disease parameters are insufficiently known. Therefore, it is still important to pursue simple methods. The simplest is lag analysis; our contribution is to evaluate leading information more quantitatively than previously attempted. Specifically, we are unaware of previous analysis with shift granularity less than one week; in contrast, our analysis tests daily shifting even if official data are less granular, and each shift is an independently computed model; thus, our ±28-day evaluation results in 57 separate models for each context.

\end{enumerate}

In summary, significant gaps remain with respect to the challenges blocking a path to an open, deployable, quantitative internet-based disease surveillance system. In this paper, we propose a path to overcoming these challenges and offer evidence demonstrating that this path is plausible.

\section{Methods}
% You may title this section "Methods" or "Models". "Models" is not a valid
% title for PLoS ONE authors. However, PLoS ONE authors may use "Analysis"

We used two data sources, Wikipedia article access logs and official disease incidence reports, and built linear models to analyze approximately 3 years of data for each of 14 disease-location contexts. This section details the nature, acquisition, and processing of these data as well as how we computed the estimation models and evaluated their output.

\subsection{Wikipedia article access logs}

% To calculate duration in days: http://www.timeanddate.com/date/duration.html
%
% ls {*/2010-03,*/2010-04,*/2010-05,*/2010-06,*/2010-07,*/2010-08,*/2010-09,*/2010-10,*/2010-11,*/2010-12,2011/*,2012/*,2013/*,*/2014-01,*/2014-02}/pagecounts-* | wc
%
% du -h --total {*/2010-03,*/2010-04,*/2010-05,*/2010-06,*/2010-07,*/2010-08,*/2010-09,*/2010-10,*/2010-11,*/2010-12,2011/*,2012/*,2013/*,*/2014-01,*/2014-02}/pagecounts-* | tail

Access logs for all Wikipedia articles are available in summary form to anyone who wishes to use them. We used the complete logs available at \url{http://dumps.wikimedia.org/other/pagecounts-raw/}. Web interfaces offering a limited view into the logs, such as \url{http://stats.grok.se}, are also available. These data are referred to using a variety of terms, including \vocab{article views}, \vocab{article visits}, \vocab{pagecount files}, \vocab{page views}, \vocab{pageviews}, \vocab{page view logs}, and \vocab{request logs}.

These summary files contain, for each hour from December 9, 2007 to present and updated in real time, a compressed text file listing the number of requests for every article in every language, except that articles with no requests are omitted. (This request count differs from the true number of human views due to automated requests, proxies, pre-fetching, people not reading the article they loaded, and other factors. However, this commonly used proxy for human views is the best available.) We analyzed data from March 7, 2010 through February 1, 2014 inclusive, a total of 1,428 days. This dataset contains roughly 34,000 data files totaling 2.7TB. 266 hours or 0.8\% of the data are missing, with the largest gap being 85 hours. These missing data were treated as zero; because they were few, this has minimal effect on our analyses.

We normalized these request counts by language. This yielded, for each article, a time series containing the number of requests for that article during each hour, expressed as a fraction of the hour's total requests for articles in the language. This normalization also compensates for periods of request undercounting, when up to 20\% fewer requests were counted than served~\cite{zachte_readme.txt_2012}. Finally, we transposed the data using Map-Reduce~\cite{dean_mapreduce:_2008} to produce files from which the request count time series of any article can be retrieved efficiently.

\subsection{Disease incidence data}

\begin{table}
  \caption{\textbf{Diseases-location contexts analyzed, with data sources.}}
  \centering
  \begin{tabular}{lllcccc}
    \hline
      \multicolumn{1}{l}{\textbf{Disease}}
    & \multicolumn{1}{l}{\textbf{Country}}
    & \multicolumn{1}{l}{\textbf{Language}}
    & \multicolumn{1}{c}{\textbf{Start}}
    & \multicolumn{1}{c}{\textbf{End}}
    & \multicolumn{1}{c}{\textbf{Resolution}}
    & \multicolumn{1}{c}{\textbf{Sources}}
    \\
    \hline
    Cholera      & Haiti         & French     & 2010-12-05 & 2013-12-05 & daily   & \cite{ministere_de_la_sante_publique_et_de_la_population_centere_????} \\
    \hline
    Dengue       & Brazil        & Portuguese & 2010-03-07 & 2013-03-16 & weekly  & \cite{ministerio_da_saude_portal_????} \\
                 & Thailand      & Thai       & 2011-01-01 & 2014-01-31 & monthly & \cite{bureau_of_epidemiology_weekly_????} \\
    \hline
    Ebola        & Uganda/DRC    & English    & 2011-01-01 & 2013-12-31 & daily   & \cite{world_health_organization_who_ebola_2011,world_health_organization_who_ebola_2012,ministere_de_la_sante_publique_republique_democratique_du_congo_fievre_2012} \\
    \hline
    HIV/AIDS     & China (PRC)   & Chinese    & 2011-01-01 & 2013-12-31 & monthly & \cite{chinese_center_for_disease_control_and_prevention_notifiable_????} \\
                 & Japan         & Japanese   & 2010-10-09 & 2013-10-18 & weekly  & \cite{national_institute_of_infectious_diseases_japan_infectious_????} \\
    \hline
    Influenza    & Japan         & Japanese   & 2010-06-26 & 2013-07-05 & weekly  & \cite{national_institute_of_infectious_diseases_japan_infectious_????} \\
                 & Poland        & Polish     & 2010-10-17 & 2013-10-23 & weekly  & \cite{national_institute_of_public_health_influenza_????} \\
                 & Thailand      & Thai       & 2011-01-23 & 2014-02-01 & weekly  & \cite{bureau_of_epidemiology_weekly_????} \\
                 & United States & English    & 2011-01-01 & 2014-01-10 & weekly  & \cite{centers_for_disease_control_and_prevention_cdc_fluview_????} \\
    \hline
    Plague       & United States & English    & 2011-01-22 & 2014-01-31 & weekly  & \cite{centers_for_disease_control_and_prevention_cdc_morbidity_????} \\
    \hline
    Tuberculosis & China (PRC)   & Chinese    & 2010-12-01 & 2013-12-31 & monthly & \cite{chinese_center_for_disease_control_and_prevention_notifiable_????} \\
                 & Norway        & Norwegian  & 2010-12-01 & 2013-12-31 & monthly & \cite{meldingssystem_for_smittsomme_sykdommer_msis_????}  \\
                 & Thailand      & Thai       & 2010-12-01 & 2013-12-31 & monthly & \cite{bureau_of_epidemiology_weekly_????} \\
    \hline
  \end{tabular}
  \begin{flushleft}
    This table lists the 7 diseases in 9 locations analyzed, for a total of 14 disease-location contexts. For each context, we list the language used as a location proxy, the inclusive start and end dates of analysis, the resolution of the disease incidence data, and one or more citations for those data.
  \end{flushleft}
  \label{tab.incidence-data}
\end{table}

Our goal was to evaluate a broad selection of diseases in a variety of countries across the world, in order to test the global applicability and disease agnosticism of our proposed technique. For example, we sought diseases with diverse modes of transmission (e.g., airborne droplet, vector, sexual, and fecal-oral), biology (virus, bacteria, protozoa), types of symptoms, length of incubation period, seasonality, and prevalence. Similarly, we sought locations in both the developed and developing world in various climates. Finally, we wanted to test each disease in multiple countries, to provide an opportunity for comparison.

These comprehensive desiderata were tempered by the realities of data availability. First, we needed reliable data establishing incidence ground truth for specific diseases in specific countries and at high temporal granularity; such official data are frequently not available for locations and diseases of interest. We used official epidemiological reports available on websites of government public health agencies as well as the World Health Organization (WHO).

Second, we needed article access counts for specific countries. This information is not present in the Wikipedia article access logs (i.e., request counts are global totals). However, a proxy is sometimes available in that certain languages are mostly limited to one country of interest; for example, a strong majority of Thai speakers are in Thailand, and the only English-speaking country where plague appears is the United States. In contrast, Spanish is spoken all over the world and thus largely unsuitable for this purpose.

Third, the language edition needs to have articles related to the disease of interest that are mature enough to evaluate and generate sufficient traffic to provide a reasonable signal.

With these constraints in mind, we used our professional judgement to select diseases and countries. The resulting list of 14 disease-location contexts, which is designed to be informative rather than comprehensive, is enumerated in Table~\ref{tab.incidence-data}.

These incidence data take two basic forms: (a)~tabular files such as spreadsheets mapping days, weeks, or months to new case counts or the total number of infected persons or (b)~graphs presenting the same mapping. In the latter case, we used plot digitizing software (Plot Digitizer, \url{http://plotdigitizer.sourceforge.net}) to extract a tabular form. We then translated these diverse tabular forms to a consistent spreadsheet format, yielding for each disease-location context a time series of disease incidence (these series are available in supplemental data~S1).

\subsection{Article selection}

The goal of our models is to create a linear mapping from the access counts of some set of Wikipedia articles to a scalar disease incidence for some disease-location context. To do so, a procedure for selecting these articles is needed; for the current proof-of-concept work, we used the following:

\begin{enumerate}

\item Examine the English-language Wikipedia article for the disease and enumerate the linked articles. Select for analysis the disease article itself along with linked articles on relevant symptoms, syndromes, pathogens, conditions, treatments, biological processes, and epidemiology. For example, the articles selected for influenza include ``Influenza'', ``Amantadine'', and ``Swine influenza'', but not ``2009 flu pandemic''.

\item Identify the corresponding article in each target language by following the inter-language wiki link; these appear at the lower left of Wikipedia articles under the heading ``Languages''. For example, the Polish articles selected for influenza include ``Grypa'',  ``Amantadyna'', and ``Świńska grypa'', but not ``Pandemia grypy A/H1N1 w latach 2009-2010'', respectively.

\item Translate each article title into the form that appears in the logs. Specifically, encode the article's Unicode title using UTF-8, percent-encode the result, and replace spaces with underscores. For example, the Polish article ``Choroby zakaźne'' becomes
\url{Choroby_zaka%C5%BAne}.
This procedure is accomplished by simply copying the article's URL from the web browser address bar.

\end{enumerate}

This procedure has two potential complications. First, an article may not exist in the target language; in this case, we simply omit it. Second, Wikipedia contains null articles called \vocab{redirects} that merely point to another article, called the \vocab{target} of the redirect. These are created to catch synonyms or common misspellings of an article. For example, in English, the article ``Flu'' is a redirect to ``Influenza''. When a user visits \url{http://en.wikipedia.org/wiki/Flu}, the content served by Wikipedia is actually that of the ``Influenza'' article; the server does not issue an HTTP 301 response nor require the reader to manually click through to the redirect target.

This complicates our analysis because this arrangement causes the redirect itself (``Flu''), not the target (``Influenza''), to appear in the access log. While in principle we could sum redirect requests into the target article's total, reliably mapping redirects to targets is a non-trivial problem because this mapping changes over time, and in fact Wikipedia's history for redirect changes is not complete~\cite{wikipedia_editors_wikipedia:moving_2014}. Therefore, we have elected to leave this issue for future work; this choice is supported by our observation below that when target and redirect are reversed, traffic to ``Dengue fever'' in Thai follows the target.

If we encounter a redirect during the above procedure, we use the target article. The complete selection of articles is available in the supplementary data~S1.

\subsection{Building and evaluating each model}

Our goal was to understand how well traffic for a simple selection of articles can nowcast and forecast disease incidence. Accordingly, we implemented the following procedure in Python to build and evaluate a model for each disease-location context.

\begin{enumerate}

\item Align the hourly article access counts with the daily, weekly, or monthly disease incidence data by summing the hourly counts for each day, week, or month in the incidence time series. This yields article and disease time series with the same frequency, making them comparable. (We ignore time zone in this procedure. Because Wikipedia data are in UTC and incidence data are in unspecified, likely local time zones, this leads to a temporal offset error of up to 23 hours, a relatively small error at the scale of our analysis. Therefore, we ignore this issue for simplicity.)

\item For each candidate article in the target language, compute Pearson's correlation $r$ against the disease incidence time series for the target country.

\item Order the candidates by decreasing $|r|$ and select the best 10 articles.

\item Use ordinary least squares to build a linear multiple regression model mapping accesses to these 10 articles to the disease time series. No other variables were incorporated into the model. Below, we report $r^2$ for the multi-article models as well as a qualitative evaluation of success or failure. We also report $r$ for individual articles in the supplementary data~S1.

\end{enumerate}

In order to test forecasting potential, we repeat the above with the article time series time-shifted from 28 days forward to 28 days backward in 1-day increments. For example, to build a 4-day forecasting model --- that is, a model that estimates disease incidence 4 days in the future --- we would shift the article time series \emph{later} by 4 days so that article request counts for a given day are matched against disease incidence 4 days in the future. The choice of ±28 days for lag analysis is based upon our \emph{a priori} hypothesis that these statistical models are likely effective for a few weeks of forecasting.

We refer to models that estimate current (i.e., same-day) disease incidence as \vocab{nowcasting} models and those that estimate past disease incidence as \vocab{anti-forecasting} models; for example, a model that estimates disease incidence 7 days ago is a 7-day anti-forecasting model. (While useless at first glance, effective anti-forecasting models that give results sooner than official data can still reduce the lead time for action. Also, it is valuable for understanding the mechanism of internet-based models to know the temporal location of predictive information.) We report $r^2$ for each time-shifted multi-article model.

\begin{table}
  \caption{\textbf{Transferability $\boldsymbol{r}_\text{t}$ example.}}
  \centering
  \begin{tabular}{lrrr}
    \hline
      \multicolumn{1}{c}{\textbf{Article}}
    & \multicolumn{1}{c}{\textbf{Japanese}}
    & \multicolumn{1}{c}{\textbf{Thai}}
    \\
    \hline
    Fever      & $ 0.23$   &  $0.21$ \\
    Chills     & $ 0.59$   &         \\
    Headache   & $-0.10$   &  $0.15$ \\
    Influenza  & $ 0.85$   &  $0.77$ \\
    \hline
  \end{tabular}
  \begin{flushleft}
    This table shows simplified models for influenza in two locations: Japan, where Japanese is spoken, and Thailand, where Thai is spoken. The Japanese model yielded correlations for Japanese versions of the articles ``Fever'', ``Chills'', ``Headache'', and ``Influenza'' of 0.23, 0.59, $-0.10$, and 0.85, respectively. The Thai model yielded correlations of 0.21, 0.15, and 0.77 for ``Fever'', ``Headache'', and ``Influenza'', respectively. Note that the article ``Chills'' is not currently present in the Thai Wikipedia. Therefore, the correlation vectors are $\{0.23, -0.10, 0.85\}$ and $\{0.21, 0.15, 0.77\}$ for the two languages. The meta-correlation, $r_\text{t}$, for these two vectors, which provides a gross estimate of how similar the models are, is 0.97. Extending this computation to the full models yields $r_\text{t} = 0.81$, as noted below in Table~\ref{tab.transferability}.
  \end{flushleft}
  \label{tab.terr-eg}
\end{table}

Finally, to evaluate whether translating models from one location to another is feasible, we compute a metric $r_\text{t}$ for each pair of locations tested on the same disease. This meta-correlation is simply the Pearson's $r$ computed between the correlation scores $r$ of each article found in both languages; the intent is to give a gross notion of similarity between models computed for the same disease in two different languages. A value of 1 means that the two models are identical, 0 means they have no relationship, and -1 means they are opposite. We ignore articles found in only one language because the goal is to obtain a sense of feasibility: given favorable conditions, could one train a model in one location and apply it to another? Table~\ref{tab.terr-eg} illustrates an example.

\section{Results}
% Results and Discussion can be combined.

Among the 14 disease-location contexts we analyzed, we found three broad classes of results. In 8 cases, the model succeeded, i.e., there was a usefully close match between the model's estimate and the official data. In 3 cases, the model failed, apparently because patterns in the official data were too subtle to capture, and in a further 3, the model failed apparently because the signal-to-noise ratio (SNR) in the Wikipedia data was too subtle to capture. Recall that this success/failure classification is based on subjective judgement; that is, in our exploration, we discovered that $r^2$ is insufficient to completely evaluate a model's goodness of fit, and a complementary qualitative evaluation was necessary.

Below, we discuss the successful and failed nowcasting models, followed by a summary and evaluation of transferability. (No models failed at nowcasting but succeeded at forecasting, so we omit a detailed forecasting discussion for brevity.)

\subsection{Successful nowcasting}

\begin{figure}
  \centering
  \includegraphics{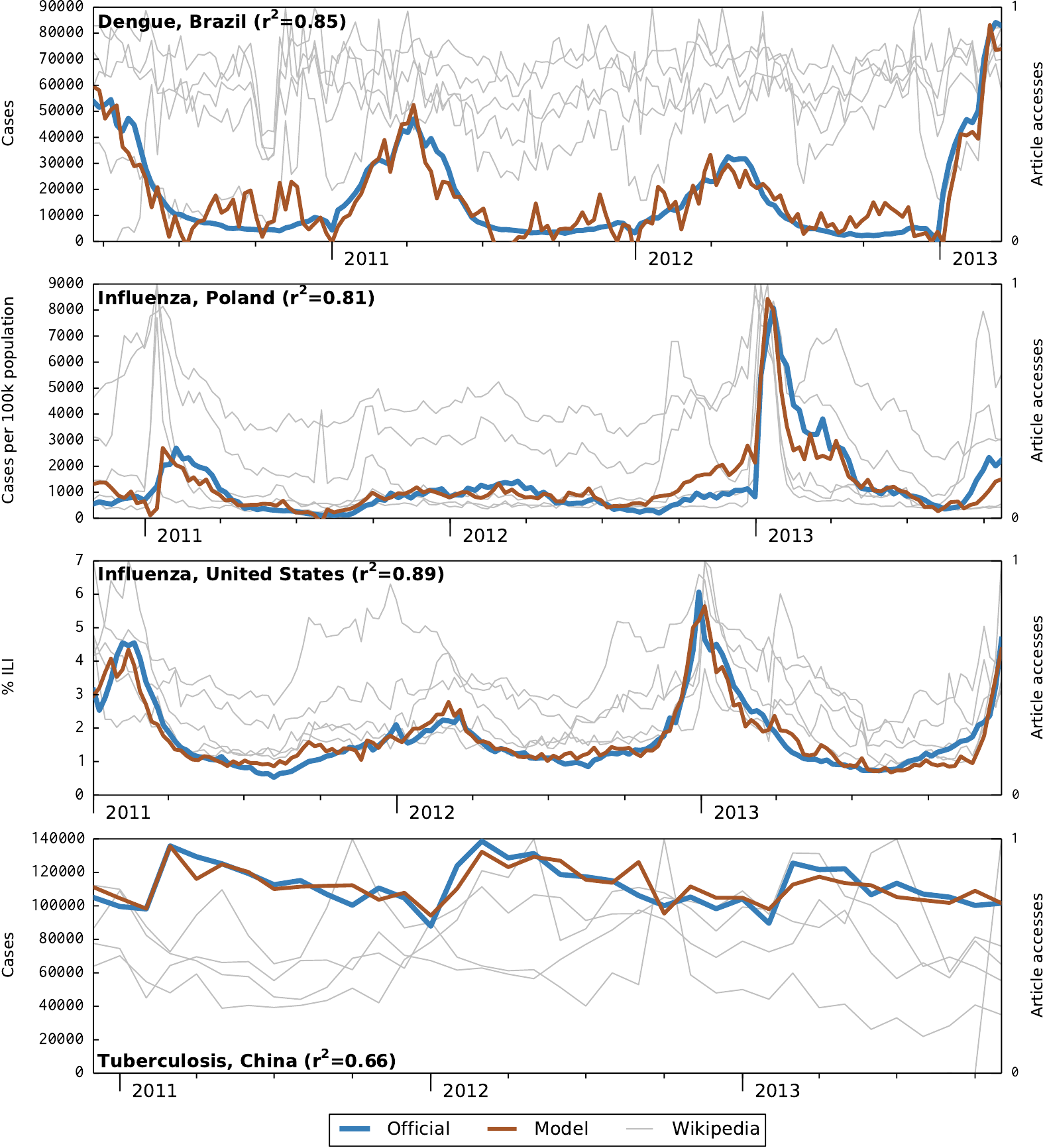}
  \caption{\textbf{Selected successful model nowcasts.} These graphs show official epidemiological data and nowcast model estimate (left Y axis) with traffic to the five most-correlated Wikipedia articles (right Y axis) over the 3 year study periods. The Wikipedia time series are individually self-normalized. Graphs for the four remaining successful contexts (dengue in Thailand, influenza in Japan, influenza in Thailand, and tuberculosis in Thailand) are included in the supplemental data file S1.}
  \label{fig.win-nowcast}
\end{figure}

% \begin{figure}
%   \centering
%   \begin{tabular}{@{}c@{}}
%     \includegraphics[width=\textwidth]{incidence_model_accesses_th_dengue_2011-01-01_2014-01-01.pdf} \\
%     \includegraphics[width=\textwidth]{incidence_model_accesses_ja_flu_2010-06-26_2013-06-29.pdf} \\
%     \includegraphics[width=\textwidth]{incidence_model_accesses_th_flu_2011-01-23_2014-01-26.pdf} \\
%     \includegraphics[width=\textwidth]{incidence_model_accesses_th_tuberculosis_2010-12-01_2013-12-01.pdf} \\
%   \end{tabular}
%   \caption{\FIXME{More graphs so we can look at them while writing. Comment out before submission.}}
%   \label{fig.win-nowcast}
% \end{figure}

Model and official data time series for selected successful contexts are illustrated in Figure~\ref{fig.win-nowcast}. The method's good performance on dengue and influenza is consistent with voluminous prior work on these diseases; this offers evidence for the feasibility of Wikipedia access as a data source.

Success in the United States is somewhat surprising. Given the widespread use of English across the globe, we expected that language would be a poor location proxy for the United States. We speculate that the good influenza model performance is due to the high levels of internet use in United States, perhaps coupled with similar flu seasons in other Northern Hemisphere countries. Similarly, in addition to Brazil, Portuguese is spoken in Portugal and several other former colonies, yet problems again failed to arise. In this case, we suspect a different explanation: the simple absence of dengue from other Portuguese-speaking countries.

The case of dengue in Brazil is further interesting because it highlights the noise inherent in this social data source, a property shared by many other internet data sources. That is, noise in the input articles is carried forward into the model's estimate. We speculate that this problem could be mitigated by building a model on a larger, more carefully selected set of articles rather than just 10.

Finally, we highlight tuberculosis in China as an example of a marginally successful model. Despite the apparently low $r^2$ of 0.66, we judged this model successful because it captured the high baseline disease level excellently and the three modest peaks well. However, it is not clear that the model provides useful information at the time scale analyzed. This result suggests that additional quantitative evaluation metrics may be needed, such as root mean squared error (RMSE) or a more complex analysis considering peaks, valleys, slope changes, and related properties.

\begin{figure}
  \centering
  \includegraphics{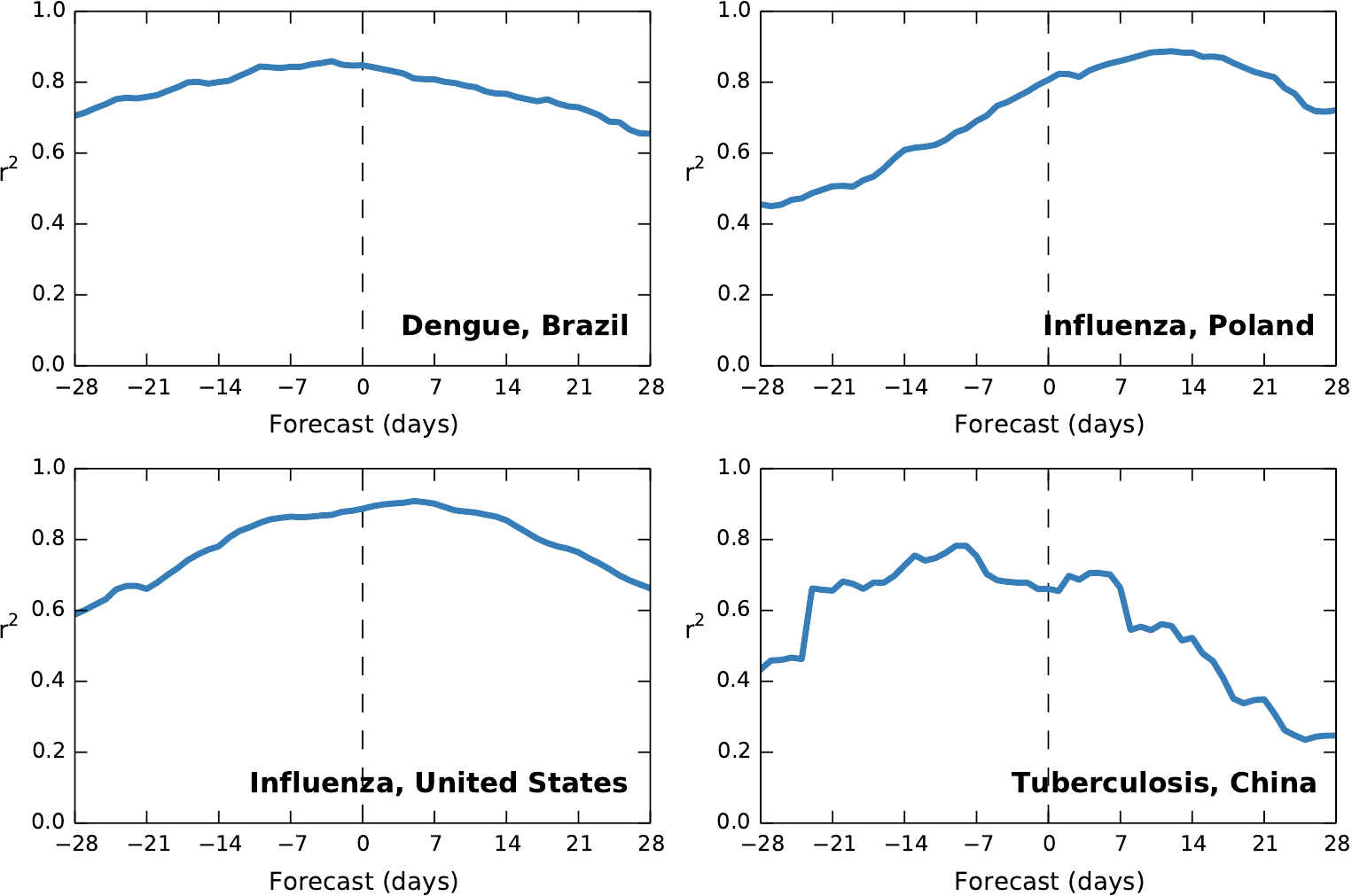}
  \caption{\textbf{Forecasting effectiveness for selected successful models.} This figure shows model $r^2$ compared to temporal offset in days: positive offsets are forecasting, zero is nowcasting (marked with a dotted line), and negative offsets are anti-forecasting. As above, figures for the four successful contexts not included here are in the supplemental data S1.}
  \label{fig.win-forecast}
\end{figure}

Forecasting and anti-forecasting performance of the four selected contexts is illustrated in Figure~\ref{fig.win-forecast}. In the case of dengue and influenza, the models contain significant forecast value through the limit of our 28-day analysis, often with the maximally effective lag comprising a forecast. We offer three possible reasons for this. First, both diseases are seasonal, so readers may simply be interested in the syndrome for this reason; however, the fact that models were able to correctly estimate seasons of widely varying severity provides counterevidence for this theory. Second, readers may be interested due to indirect reasons such as news coverage. Prior work disagrees on the impact of such influences; for example, Dukic~\etal\ found that adding news coverage to their methicillin-resistant \emph{Staphylococcus aureus} (MRSA) model had a limited effect~\cite{dukic_internet_2011}, but recent Google Flu Trends failures appear to be caused in part by media activity~\cite{butler_when_2013}. Finally, both diseases have a relatively short incubation period (influenza at 1--4 days and dengue at 3--14); soon-to-be-ill readers may be observing the illness of their infectors or those who are a small number of degrees removed. It is the third hypothesis that is most interesting for forecasting purposes, and evidence to distinguish among them might be obtained from studies using simulated media and internet data, as suggested by Culotta~\cite{culotta_lightweight_2013}.

Tuberculosis in China is another story. In this case, the model's effectiveness is poorer as the forecast interval increases; we speculate that this is because seasonality is absent and the incubation period of 2--12 weeks is longer, diluting the effect of the above two mechanisms.

\subsection{Failed nowcasting}

\begin{figure}
  \centering
  \includegraphics{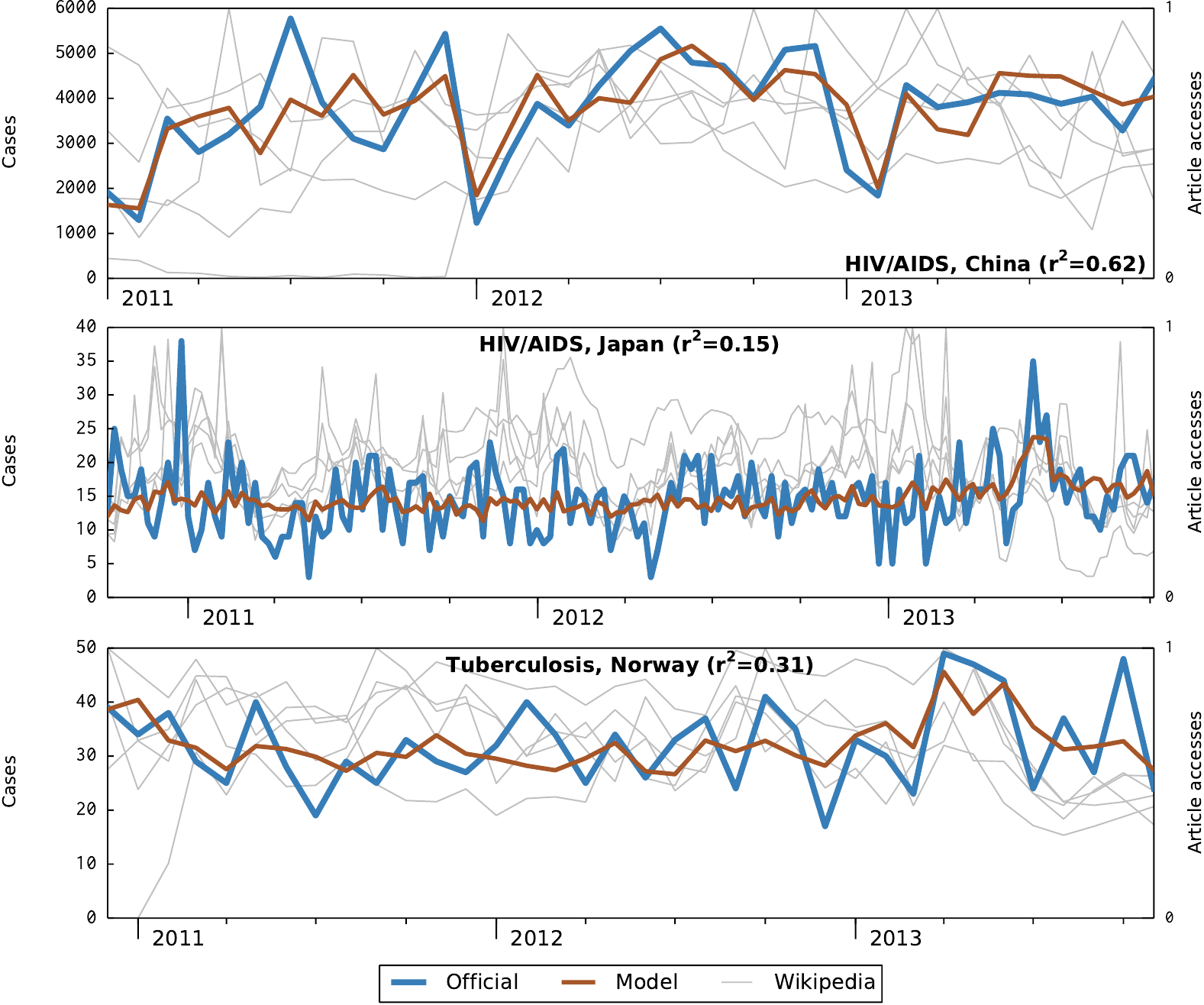}
  \caption{\textbf{Nowcast attempts where the model was unable to capture a meaningful pattern in official data.}}
  \label{fig.fail-variation}
\end{figure}

Figure~\ref{fig.fail-variation} illustrates the three contexts where the model was not effective because, we suspect, it was not able to discern meaningful patterns in the official data. These suggest a few patterns that models might have difficulty with:

\begin{enumerate}

\item \inhead{Noise.} True patterns in data may be obscured by noise. For example, in the case of HIV/AIDS in China, the official data vary by a factor of 2 or more throughout the graph, and the model captures this fairly well, but the pattern seems epidemiologically strange and thus we suspect it may be merely noise. The other two contexts appear to also contain significant noise.

(Note that we distinguish noisy official data from an unfavorable signal-to-noise ratio, which is discussed below.)

\item \inhead{Too slow.} Disease incidence may be changing too slowly to be evident in the chosen analysis period. In all three contexts shown in Figure~\ref{fig.fail-variation}, the trend of the official data is essentially flat, with HIV/AIDS in Japan especially so. The models have captured this flat trend fairly well, but even doing so excellently provides little actionable value over traditional surveillance.

Both HIV/AIDS and tuberculosis infections progress quite slowly. A period of analysis longer than three years might reveal meaningful patterns that could be captured by this class of models. However, the social internet is young and turbulent; for example, even 3 years consumes most of the active life of some languages of Wikipedia. This complicates longitudinal analyses.

\item \inhead{Too fast.} Finally, incidence may be changing too quickly for the model to capture. We did not identify this case in the contexts we tested; however, it is clearly plausible. For example, quarterly influenza data would be hard to model meaningfully using these techniques.

\end{enumerate}

In all three patterns, improvements such as non-linear models or better regression techniques could lead to better results, suggesting that this is a useful direction for future work. In particular, noise suppression techniques as well as models tuned for the expected variation in a particular disease may prove fruitful.

\begin{figure}
  \centering
  \includegraphics{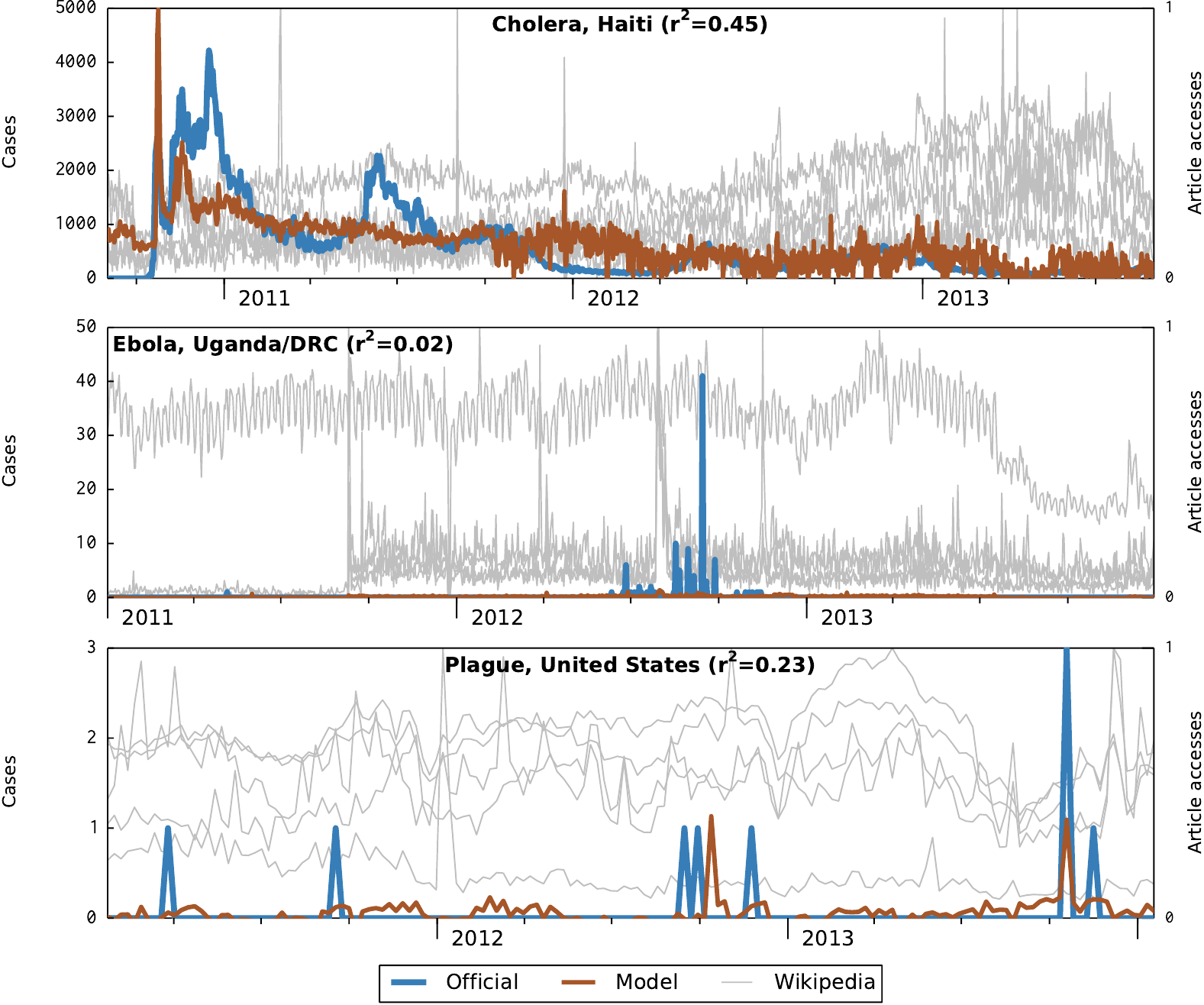}
  \caption{\textbf{Nowcast attempts with poor performance due to unfavorable signal-to-noise ratio.}}
  \label{fig.fail-signal}
\end{figure}

Figure~\ref{fig.fail-signal} illustrates the three contexts where we suspect the model failed due to a signal-to-noise ratio (SNR) in the Wikipedia data that was too low. That is, the number of Wikipedia accesses due to actual observations of infection is drowned out by accesses due to other causes.

In the case of Ebola, there are relatively few direct observations (a major outbreak has tens of cases), and the path to these observations becoming internet traces is hampered by poor connectivity in the sub-Saharan countries where the disease is active. On the other hand, the disease is one of general ongoing interest; in fact, one can observe on the graph a pattern of weekly variation (higher in midweek, lower on the weekend), which is common in online activity. In combination, these yield a completely useless model.

The United States has good internet connectivity, but plague has even lower incidence (the peak on the graph is three cases) and this disease is also popularly interesting, resulting in essentially the same effect. The cholera outbreak in Haiti differs in that the number of cases is quite large (the peak of the graph is 4,200 cases in one day). However, internet connectivity in Haiti was already poor even before the earthquake, and the outbreak was a major world news story, increasing noise, so the signal was again lost.

%\begin{figure}
%  \centering
%\includegraphics[width=\textwidth]{incidence_model_accesses_fr_cholera_2010-10-10_2010-10-31.pdf}
%  \caption{\textbf{Model nowcast during the emergence of the 2010 cholera outbreak in Haiti.}}
%  \label{fig.cholera}
%\end{figure}

% See issue #71 for why this is commented out.
%
% In fact, Figure~\ref{fig.cholera} makes it clear that the model's estimate is following the news rather than incidence; this suggests that prior successes measuring this outbreak with social internet data should be interpreted with caution.

\subsection{Performance summary}

\begin{table}
  \caption{\textbf{Model performance summary}}
  \centering

\begin{tabular}{|ll|l|rrrr|rr|}
\hline

&
&
& \multicolumn{4}{c|}{\textbf{$\boldsymbol{r^2}$ at forecast}}
& \multicolumn{2}{c|}{\textbf{Best forec.}}
\\
  \multicolumn{1}{|c}{\textbf{Disease}}
& \multicolumn{1}{c}{\textbf{Location}}
& \multicolumn{1}{|c|}{\textbf{Result}}
& \multicolumn{1}{c}{\textbf{0}}
& \multicolumn{1}{c}{\textbf{7}}
& \multicolumn{1}{c}{\textbf{14}}
& \multicolumn{1}{c|}{\textbf{28}}
& \multicolumn{1}{c}{\textbf{Days}}
& \multicolumn{1}{c|}{$\boldsymbol{r^2}$}
\\

\hline
Cholera & Haiti & Failure (SNR) & 0.45 & 0.39 & 0.41 & 0.48 & 26 & 0.50 \\
\hline
Dengue & Brazil & Success & 0.85 & 0.81 & 0.77 & 0.65 & -3 & 0.86 \\
 & Thailand & Success & 0.55 & 0.54 & 0.57 & 0.74 & 28 & 0.74 \\
\hline
Ebola & Uganda/DRC & Failure (SNR) & 0.02 & 0.01 & 0.02 & 0.02 & 5 & 0.14 \\
\hline
HIV/AIDS & China (PRC) & Failure (Official data) & 0.62 & 0.48 & 0.34 & 0.31 & -1 & 0.63 \\
 & Japan & Failure (Official data) & 0.15 & 0.19 & 0.15 & 0.05 & 9 & 0.22 \\
\hline
Influenza & Japan & Success & 0.82 & 0.92 & 0.86 & 0.52 & 8 & 0.92 \\
 & Poland & Success & 0.81 & 0.86 & 0.88 & 0.72 & 12 & 0.89 \\
 & Thailand & Success & 0.79 & 0.76 & 0.67 & 0.48 & -2 & 0.80 \\
 & United States & Success & 0.89 & 0.90 & 0.85 & 0.66 & 5 & 0.91 \\
\hline
Plague & United States & Failure (SNR) & 0.23 & 0.03 & 0.05 & 0.07 & 0 & 0.23 \\
\hline
Tuberculosis & China (PRC) & Success & 0.66 & 0.66 & 0.52 & 0.25 & -9 & 0.78 \\
 & Norway & Failure (Official data) & 0.31 & 0.41 & 0.40 & 0.42 & 20 & 0.48 \\
 & Thailand & Success & 0.68 & 0.68 & 0.69 & 0.69 & 9 & 0.69 \\

\hline
\end{tabular}

  \begin{flushleft}
    This table summarizes the performance of our estimation models. For each disease and location, we list the subjective success/failure classification as well as model $r^2$ at nowcasting (0-day forecast) and 7-, 14-, and 28-day forecasts. We also list the temporal offset in days of the best model (again, a positive offset indicates forecasting) along with that model's $r^2$.
  \end{flushleft}
  \label{tab.summary}
\end{table}

Table~\ref{tab.summary} summarizes the performance of our models in the 14 disease-location contexts tested. Of these, we classified 8 as successful, producing useful estimates for both nowcasting and forecasting, and 6 as unsuccessful. Performance roughly broke down along disease lines: all influenza and dengue models were successful, while two of the three tuberculosis models were, and cholera, ebola, HIV/AIDS, and plague proved unsuccessful. Given the relatively simple model building technique used, this suggests that our Wikipedia-based approach is sufficiently promising to explore in more detail. (Another hypothesis is that model performance is related to popularity of the corresponding Wikipedia language edition. However, we found no relationship between $r^2$ and either a language's total number of articles or total traffic.)

At a higher level, we posit that a successful estimation model based on Wikipedia access logs or other social internet data requires two key elements. First, it must be sensitive enough to capture the true variation in disease incidence data. Second, it must be sensitive enough to distinguish between activity traces due to health-related observations and those due to other causes. In both cases, further research on modeling techniques is likely to yield sensitivity improvements. In particular, a broader article selection procedure --- for example, using big data methods to test \textit{all} non-trivial article time series for correlation, as Ginsberg~\etal\ did for search queries~\cite{ginsberg_detecting_2008} --- is likely to prove fruitful, as might a non-linear statistical mapping.

\subsection{Transferability}

\begin{table}
  \caption{\textbf{Transferability scores $\boldsymbol{r}_\text{t}$ for paired models}}
  \centering
  \begin{tabular}{|lllr|}
    \hline
      \multicolumn{1}{|c}{\textbf{Disease}}
    & \multicolumn{1}{c}{\textbf{Location 1}}
    & \multicolumn{1}{c}{\textbf{Location 2}}
    & \multicolumn{1}{c|}{$\boldsymbol{r_\textbf{t}}$} \\
    \hline
    Dengue       & Brazil      & Thailand      &  0.39 \\
    \hline
    HIV/AIDS     & China (PRC) & Japan         & -0.06 \\
    \hline
    Influenza    & Japan       & Poland        &  0.45 \\
                 & Japan       & Thailand      &  0.81 \\
                 & Japan       & United States &  0.62 \\
                 & Poland      & Thailand      &  0.48 \\
                 & Poland      & United States &  0.44 \\
                 & Thailand    & United States &  0.76 \\
    \hline
    Tuberculosis & China (PRC) & Norway        &  0.19 \\
                 & China (PRC) & Thailand      & -0.20 \\
                 & Norway      & Thailand      &   n/a \\
    \hline
  \end{tabular}
  \begin{flushleft}
    This table lists the transferability scores $r_\text{t}$ for each tested pair of countries within a disease. Countries that did not share enough articles to compute a meaningful $r_\text{t}$ are indicated with \textit{n/a}.
  \end{flushleft}
  \label{tab.transferability}
\end{table}

Table~\ref{tab.transferability} lists the transferability scores $r_\text{t}$ for each pair of countries tested on the same disease. Because this paper is concerned with establishing feasibility, we focus on the highest scores. These are encouraging: in the case of influenza, both Japan/Thailand and Thailand/United States are promising. That is, it seems plausible that careful source model selection and training techniques may yield useful models in contexts where no training data are available (e.g., official data are unavailable or unreliable). These early results suggest that research to quantitatively test methods for translating models from one disease-location context to another should be pursued.

\section{Discussion}

Human activity on the internet leaves voluminous traces that contain real and useful evidence of disease dynamics. Above, we pose four challenges currently preventing these traces from informing operational disease surveillance activities, and we argue that Wikipedia data are one of the few social internet data sources that can meet all four challenges. Specifically:

\begin{enumerate}

\item[C1.] \inhead{Openness.} Open data and algorithms are required, in order to offer reliable science as well as a flexible and robust operational capability. Wikipedia access logs are freely available to anyone.

\item[C2.] \inhead{Breadth.} Thousands of disease-location contexts, not dozens, are needed to fully understand the global disease threat. We tested simple disease estimation models on 14 contexts around the world; in 8 of these, the models were successful with $r^2$ up to 0.92, suggesting that Wikipedia data are useful in this regard.

\item[C3.] \inhead{Transferability.} The greatest promise of novel disease surveillance methods is the possibility of use in contexts where traditional surveillance is poor or nonexistent. Our analysis uncovered pairs of same-disease, different-location models with similarity up to 0.81, suggesting that translation of trained models using Wikipedia's mappings of one language to another may be possible.

\item[C4.] \inhead{Forecasting.} Effective response to disease depends on knowing not only what is happening now but also what will happen in the future. Traditional mechanistic forecasting models often cannot be applied due to missing parameters, motivating the use of simpler statistical models. We show that such statistical models based on Wikipedia data have forecasting value through our maximum tested horizon of 28 days.

\end{enumerate}

This preliminary study has several important limitations. These comprise an agenda for future research work:

\begin{enumerate}

\item The methods need to be tested in many more contexts in order to draw lessons about when and why this class of methods is likely to work.

\item A better article selection procedure is needed. In the current paper, we tried a simple manual process yielding at most a few dozen candidate articles in order to establish feasibility. However, real techniques should use a comprehensive process that evaluates thousands, millions, or all plausible articles for inclusion in the model. This will also facilitate content analysis studies that evaluate which types of articles are predictive of disease incidence.

\item Better geo-location is needed. While language as a location proxy works well in some cases, as we have demonstrated, it is inherently weak. In particular, it is implausible for use at a finer scale than country-level. What is needed is a hierarchical geographic aggregation of article traffic. The Wikimedia Foundation, operators of Wikipedia and several related projects, could do this using IP addresses to infer location before the aggregated data are released to the public. For most epidemiologically-useful granularities, this will still preserve reader privacy.

\item Statistical estimation maps from article traffic to disease incidence should be more sophisticated. Here, we tried simple linear models mapping a single interval's Wikipedia traffic to a single interval's disease incidence. Future directions include testing non-linear and multi-interval models.

\item Wikipedia data have a variety of instabilities that need to be understood and compensated for. For example, Wikipedia shares many of the problems of other internet data, such as highly variable interest-driven traffic caused by news reporting and other sources.

  Wikipedia has its own data peculiarities that can also cause difficulty. For example, during preliminary exploration for this paper in spring 2013, we used the inter-language link on the English article ``Dengue fever'' to locate the Thai version, ``\raisebox{-1pt}{\includegraphics{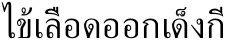}}'' (roughly, ``dengue hemorrhagic fever''); article access logs indicated several hundred accesses per day for this article in the month of June 2013. When we repeated the same process in March 2014, the inter-language link led to a page with the same content, but a different title, ``\raisebox{-1pt}{\includegraphics{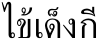}}'' (roughly, ``dengue fever''). As none of the authors are Thai speakers, and Google Translate renders both versions as ``dengue fever'', we did not notice that the title of the Thai article had changed and were alarmed to discover that the article's traffic in June 2013 was essentially zero.

  The explanation is that before July 23, 2013, ``\raisebox{-1pt}{\includegraphics{thaishort.pdf}}'' was a redirect to ``\raisebox{-1pt}{\includegraphics{thailong.pdf}}''; on that day, the direction of the redirect was reversed, and almost all accesses moved over to the new redirect target over a period of a few days. That is, the article was the same all along, but the URL under which its accesses were recorded changed.

  Possible techniques for compensation include article selection procedures that exclude such articles or maintaining a time-aware redirect graph so that different aliases of the same article can be merged. Indeed, when we tried the latter approach by manually summing the two URLs' time series, it improved nowcast $r^2$ from 0.55 to 0.65. However, the first technique is likely to discard useful information, and the second may not be reliable because complete history for this type of article transformation is not available~\cite{wikipedia_editors_wikipedia:moving_2014}.

  In general, ongoing, time-aware re-training of models will likely be helpful, and limitations of the compensation techniques can be evaluated with simulation studies that inject data problems.

\item We have not explored the full richness of the Wikipedia data. For example, complete histories of each language edition are available, which include editing metadata (timestamps, editor identity, and comments), the text of each version, and conversations about the articles; these would facilitate analysis of edit activity as well as the articles' changing text. Also, health-related articles are often mapped to existing ontologies such as the International Statistical Classification of Diseases and Related Health Problems (ICD-9 or ICD-10).

\item Transferability of models should be tested using more realistic techniques, such as simply building a model in one context and testing its performance in another.

\end{enumerate}

Finally, it is important to recognize the biases inherent in Wikipedia and other social internet data sources. Most importantly, the data strongly over-represent people and places with good internet access and technology skills; demographic biases such as age, gender, and education also play a role. These biases are sometimes quantified (e.g., with survey results) and sometimes completely unknown. As noted above, simulation studies using synthetic internet data can quantify the impact and limitations of these biases.

Despite these limitations, we have established the utility of Wikipedia access logs for global disease monitoring and forecasting, and we have outlined a plausible path to a reliable, scientifically sound, operational disease surveillance system. We look forward to collaborating with the scientific and technical community to make this vision a reality.

\section{Acknowledgments}

Mac Brown's participation in discussions improved this work significantly. This work is supported in part by NIH/NIGMS/MIDAS under grant U01-GM097658-01 and the Defense Threat Reduction Agency (DTRA), Joint Science and Technology Office for Chemical and Biological Defense under project numbers CB3656 and CB10007. Data collected using QUAC; this functionality was supported by the U.S.\ Department of Energy through the LANL LDRD Program. Computation used HPC resources provided by the LANL Institutional Computing Program. LANL is operated by Los Alamos National Security, LLC for the Department of Energy under contract DE-AC52-06NA25396. Approved for public release: LA-UR~14-22535.

\section{Supplementary information}

\paragraph{Data S1} \textbf{Input data, raw results, and additional figures.} This archive file contains: (a)~inter-language article mappings, (b)~figures for the 4 successful contexts not included above, (c)~official epidemiological data used as input, (d)~complete correlation scores $r$, (e)~wiki input data, and (f)~a text file explaining the archive content and file formats.

%\section*{References}

% How to update the references:
%
% 1. Export everything in the paper collection to refs-raw.bib.
% 2. Find the new or updated reference in that file.
% 3. Add reference to refs-cooked.bib (be sure to keep the same order).
%    a. git diff refs-raw.bib | patch refs-cooked.bib
% 4. Make necessary corrections:
%    a. Remove address and other cruft that seems extraneous.
%    b. Clean up title capitalization.
%    d. Clean up journal or booktitle.
%    e. Prepend "Proc.\ " to conference booktitles.
%    e. Add sort key if needed.
%
% Note: Some of the above may be better done in Zotero. Use your judgement.

\bibliography{refs-cooked}

%\section{Figure Legends}
%\section{Tables}

\end{document}